\documentclass{aastex}
\usepackage{spr-astr-addons}
\usepackage{url}

\RequirePackage{color}

\newcommand{\beq}{\begin{equation}}
\newcommand{\eeq}{\vspace{0cm} \end{equation}}
\newcommand{\beqq}{\setlength\arraycolsep{2pt}\begin{eqnarray}}
\newcommand{\eeqq}{\vspace{0cm} \end{eqnarray}}

\begin{document}

\title{Thermodynamics of Chaplygin gas}

\shorttitle{Thermodynamics of Chaplygin gas} \shortauthors{Myung}

\author{Yun Soo Myung}

\affil{Institute of Basic Science and School of
    Computer Aided Science,  Inje University, Gimhae 621-749, Korea.\\ Email: ysmyung@inje.ac.kr}

\begin{abstract}
We clarify   thermodynamics of the Chaplygin gas by introducing the
integrability condition. All thermal quantities  are derived as
functions of either volume or temperature. Importantly, we find a
new general equation of state, describing the Chaplygin gas
completely. We confirm that the Chaplygin gas could show  a unified
picture of dark matter and energy which  cools down through the
universe expansion without any critical point (phase transition).
\end{abstract}

\section{Introduction}
We start with an exotic perfect fluid, named Chaplygin gas which
obeys the following adiabatic equation of state \citep{KMP,FGS}
\begin{equation} \label{ceos}
p=-\frac{A}{\rho^{\alpha}},
\end{equation}
where $\rho$ is the energy density of the fluid defined by
$\rho=E/V$ and $\alpha$ is constant and positive: $\alpha>0$. In
this work, we choose $\alpha=1$ as the Chaplygin gas for simplicity.
The parameter $A$ is positive and considered as a universal
constant. Here we choose $A=1$ for numerical computations. This gas
behaves as pressureless gas (dust) at high energy densities, while
it behaves a cosmological constant with negative pressure at low
energy densities. Hence the Chaplygin gas is regarded as a unified
model of dark matter and dark energy \citep{BBS,BBT,FGVZ}.
 For the thermodynamic study \citep{SBS,SSB}, we apply the combination of the
first- and second-law of thermodynamics to the system with volume
$V$. Then it leads to
 \begin{equation} TdS=d(\rho V)+
p dV = d[(\rho+p)V]-V dp \label{kt364}. \end{equation} The
integrability condition is necessary to define the Chaplygin gas  as
a thermodynamic system~\citep{KT,GWW,MyungP}. It is given by \beq
\frac{\partial^2 S}{\partial T \partial V}=\frac{\partial^2
S}{\partial V \partial T}\label{kt365} \eeq which leads to the
 relation between the pressure (energy density) and temperature  \beq
 dp=\frac{\rho+p}{
T}dT.\label{kt367} \eeq Plugging Eq.(\ref{kt367}) into
Eq.(\ref{kt364}), we have the differential relation, \beq dS =
\frac{1}{ T} d[(\rho + p)V]-(\rho + p)V \frac{dT}{
T^2}=d\Bigg[\frac{(\rho+p)V}{ T} + C \Bigg]\label{kt368} \eeq where
$C$ is a constant. The entropy  is  defined by \beq S \equiv
\frac{(\rho + p)}{ T}V\label{entropy} \eeq up to an additive
constant. Even for an adiabatic process of $S={\rm const}$, the same
definition of entropy follows from the conservation law which can be
rewritten as \beq
 d[(\rho+p)V]=V dp.\label{kt369} \eeq Inserting the integrability condition Eq.(\ref{kt367}) into Eq.(\ref{kt369}), one
recovers Eq.(\ref{entropy}) immediately. Hence we use the equation
(\ref{entropy}) as a defining equation of the temperature for an
adiabatic process~\citep{LA,GS} \beq T \equiv \frac{(\rho + p)}{
S}V. \label{temp} \eeq On the other hand, the conservation law
(\ref{kt369}) plays an important role in
 the homogeneous and
isotropic FRW universe which is
 described by two Friedmann equations based on the
Robertson-Walker metric \beqq  \label{1stf}
H^2&=& \frac{8 \pi G}{3}\rho-\frac{k}{a^2},\\
  \label{secf} \dot{H}&=&-4\pi G(\rho+p)+\frac{k}{a^2} \eeqq
where $H=\dot{a}/a$ is the Hubble parameter and $k=-1,0,1$ represent
the three-dimensional space with the negative, zero, and positive
spatial curvature, respectively. The conservation-law could be
derived from Eqs.(\ref{1stf}) and (\ref{secf}) as \beq
\label{con-law} \dot{\rho}+3H(\rho+p)=0 \eeq
 Hence, one
equation among three is redundant. Here we may choose  the first
Friedmann equation (\ref{1stf})  and the conservation law
(\ref{con-law}) as two relevant equations. Importantly,
Eq.(\ref{con-law}) together with Eq.(\ref{ceos}) is solved to give
the energy density
 \beq
\label{rho-a}\rho(a)=\sqrt{A+\frac{B}{a^6}},\eeq  where $B$ is an
integration constant. By choosing a positive value for $B$, we see
that for small $a$ $(a^6 \ll B/A)$, Eq.(\ref{rho-a}) and the
pressure take the forms approximately \beq \label{rho-app}\rho \sim
\frac{\sqrt{B}}{a^3},~p \sim 0\eeq which corresponds to a
matter-dominated universe. For a large value $a$, it follows that
\beq \label{rho-la}\rho \sim \sqrt{A},~~p\sim -\sqrt{A} \eeq which
corresponds to a dark energy-dominated universe.

Taking $V=a^3$, we have the relations \beq
\rho(V)=\sqrt{A+\frac{B}{V^2}},~p(V)=-\frac{A}{\sqrt{A+\frac{B}{V^2}}}.
\eeq Considering Eq.(\ref{temp}), one has the temperature\beq T(V)=
\frac{1}{S\sqrt{AV^2+B}}. \eeq Furthermore, the equation of state
$\omega(V)$  and squared speed of sound $v^2(V)$ \citep{Myungh} is
given by \beq \omega(V)\equiv
\frac{p}{\rho}=-\frac{A}{A+\frac{B}{V^2}},~v^2(V)\equiv
\frac{\partial p}{\partial \rho}=\frac{A}{A+\frac{B}{V^2}}. \eeq In
this case there is no restriction on thermodynamic quantities as
functions of $V$ ($0\le V\le \infty$). As is shown in Fig. 1, we
recover the matter-dominated universe of $\rho  \to \infty,~p\to 0$
as $V\to 0$, while the dark-energy dominated universe of $\rho  \to
1,~p\to -1$  appears as $V \to \infty$. Also we  note that the
temperature goes zero only when the volume goes to infinity, showing
that the third-law of thermodynamics is satisfied with the Chaplygin
gas, contrast to Ref.\citep{SBS}.
\begin{figure}[t!]
   \centering
   \includegraphics[scale=.8]{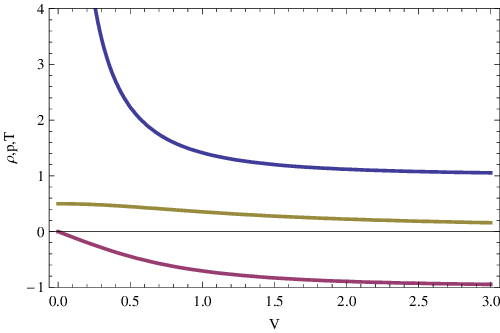}
\caption{Graphs of energy density $\rho$ (top), pressure $p$
(bottom), and temperature $T$(middle) functions of $V$ with $A=B=1$
and $S=2$. } \label{fig1}
\end{figure}

\section{Thermodynamics}
 From now on, we wish to derive two important relations
 from the integrability condition Eq.(\ref{kt367}).  We
have one connection for the pressure and temperature \beq
\frac{p~dp}{p^2-A}= \frac{dT}{T} \eeq which leads to \beq \ln
\Big[\frac{p^2-A}{-A}\Big]=\ln\Big[\frac{T^2}{T^2_*}\Big]. \eeq Here
$T_*$ is the temperature corresponding to $p=0$.  In this case, we
obtain the relation between pressure and temperature~\citep{SBS}
 \beq
p(T)=-\sqrt{A}\sqrt{1-\frac{T^2}{T^2_*}}. \label{pt-law}\eeq
 Also, from the other
form of integrability condition \beq \frac{Ad\rho}{\rho(\rho^2-A)}=
\frac{dT}{T}, \eeq we derive the important relation~\citep{SBS} \beq
\label{sb-law} \rho(T)=\frac{\sqrt{A}}{\sqrt{1-\frac{T^2}{T^2_*}}}.
\eeq Here we observe the allowed range of temperature: $0\le T \le
T_{*}$. Using  these expressions, one has two quantities of equation
of state  and squared speed of sound  as function of $T$ \beq
\omega(T)=-\frac{A}{\rho^2(T)}=-1+\frac{T^2}{T^2_*},~v^2(T)=\frac{\partial
p}{\partial \rho}=1-\frac{T^2}{T^2_*}
 \eeq
which are  important to describe a fluid of the  Chaplygin gas.

 Finally, the heat capacity is calculated to be \beq
C_V(T)=V \frac{\partial \rho}{\partial T}= \frac{\sqrt{A}
V}{T^2_*}\frac{T}{\Big[1-\frac{T^2}{T^2_*}\Big]^{3/2}}. \eeq
 All pictures of thermodynamic quantities are depicted as functions of
 temperature in Fig. 2. These show that an evolution  from $T=T^*=100$ to $T=0$ is
a thermodynamically stable transition without any critical point.
That is, even though the Chaplygin gas provides an accelerating
universe, its fluid has a stable nature of positive squared sound
velocity ($v^2=-\omega^2$) and positive heat capacity during the
evolution.

In this section we recover all known thermodynamic quantities as
functions of temperature using the integrability condition .
\begin{figure}[t!]
   \centering
   \includegraphics[scale=.7]{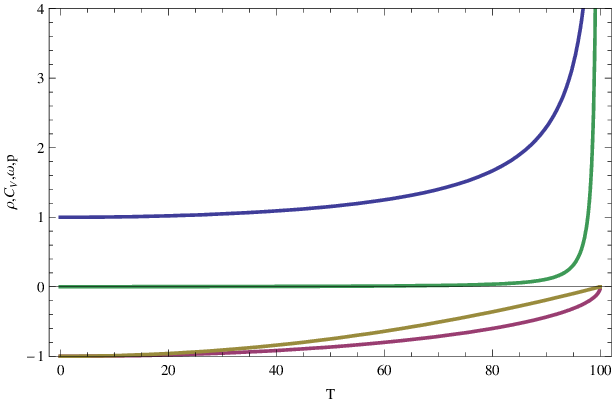}
\caption{Graphs of energy density $\rho(T)$,  heat capacity
$C_V(T)$,  equation of state $\omega(T)$, and pressure $p(T)$ from
top to bottom with $A=1$ and $V=1$. All these show that a process
from $T=T^*=100$ to $T=0$ is a thermodynamically stable transition.
} \label{fig2}
\end{figure}

\section{ General equation of state}

Combining Eq.(\ref{temp}) with Eq.(\ref{ceos}), we have the
second-order equation for pressure $p$
 \beq Vp^2-TSp -AV=0, \eeq
whose solution leads to  a complete equation of state for $p,V,T$
\beq  \label{pvt}pV=T\Bigg[\frac{S}{2}-\sqrt{A
\frac{V^2}{T^2}+\frac{S^2}{4}}\Bigg]. \eeq For the case of $S=2$, it
takes a simpler form \beq \label{spvt}pV=T\Bigg[1-\sqrt{1+A
\frac{V^2}{T^2}}\Bigg]. \eeq As far as we know, this is the first
equation of state for an adiabatic Chaplygin gas. For an ideal gas,
we have $pV^{\gamma}={\rm const}$ with $\gamma=C_p/C_V$.

 On the other hand, we have the
second-order equation for energy density $\rho$
 \beq V\rho^2-TS\rho -AV=0, \eeq
whose solution leads to \beq  \label{rvt} \rho
V=T\Bigg[\frac{S}{2}+\sqrt{A \frac{V^2}{T^2}+\frac{S^2}{4}}\Bigg].
\eeq For the case of $S=2$, it takes a simpler form \beq
\label{srvt}\rho V=T\Bigg[1+\sqrt{1+A \frac{V^2}{T^2}}\Bigg]. \eeq
We check that adding (\ref{pvt}) and (\ref{rvt}) leads to
(\ref{temp}). Also dividing (\ref{pvt}) by (\ref{rvt}) leads to the
equation of state as functions of $T$ and $V$ \beq
\omega(T,V)=-\frac{A}{\rho(T,V)}. \eeq

We consider the case of a constant temperature (See Figs. 1 and 3).
In the limit of $V\to \infty$, one has \beq \rho \to \sqrt{A},~p \to
-\sqrt{A}, \eeq while in the limit of $ V \to 0$, one finds \beq
\rho \to \frac{2T}{V}(=\infty),~p \to 0. \eeq Comparing Fig. 3 with
Fig. 1, we find that Eqs.(\ref{spvt}) and (\ref{srvt}) could
describe an adiabatic, isothermal process of the Chaplygin gas well.

 For the case of
constant volume (See Figs. 2 and 4), one has \beq \rho \to
\sqrt{A},~p \to -\sqrt{A} \eeq in the limit of $T\to0$, while in the
limit of $ T\to \infty$, one finds \beq\rho \to
\frac{2T}{V}(=\infty),~ p \to 0. \eeq The former describes the dark
energy-dominated universe, while the latter describes the
matter-dominated universe. We find that Eqs.(\ref{spvt}) and
(\ref{srvt}) could describe an adiabatic, isochoric process of the
Chaplygin gas well.
\begin{figure}[t!]
   \centering
   \includegraphics[scale=.9]{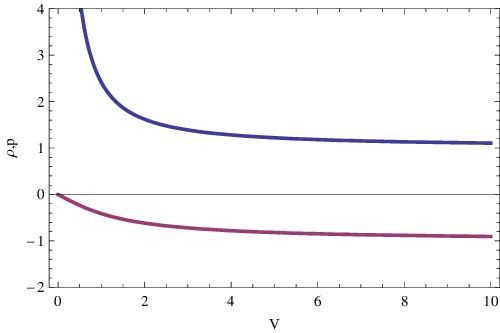}
\caption{Graphs of $\rho(T=1,V)$ and  $p(T=1,V)$ from top to bottom
with $A=1$ and $S=2$. This mimics Fig. 1. } \label{fig3}
\end{figure}
\begin{figure}[t!]
   \centering
   \includegraphics[scale=.9]{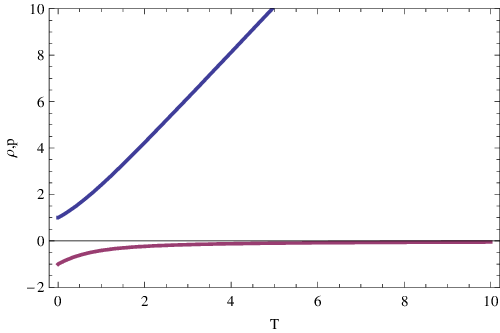}
\caption{Graphs of  $\rho(T,V=1)$ and $p(T,V=1)$ from top to bottom
with $A=1$ and $S=2$. This mimics Fig. 2. } \label{fig4}
\end{figure}

\section{Discussions}
We clarify  thermodynamics of the Chaplygin gas by introducing the
integrability condition Eq. (\ref{kt365})  and the temperature of
(\ref{temp}).  All thermal quantities are derived as functions of
either temperature or volume. In this case, we show that the
third-law of thermodynamics is satisfied with the Chaplygin gas.
Furthermore, we find a new general  equation of state, describing
the Chaplygin gas as function of temperature and volume completely.
For the generalized Chaplygin gas with $\alpha>0$, we expect to have
similar behaviors as the Chaplygin gas did show.

Consequently, we confirm that the Chaplygin gas could show a unified
picture of dark matter and energy which cools down through the
universe expansion without any critical point.

\begin{acknowledgments}
 This work was supported by the National Research
Foundation of Korea (NRF) grant funded by the Korea government
(MEST) \\
 (No.2010$-$0028080).
\end{acknowledgments}

\end{document}